\documentstyle[12pt,epsfig,latexsym]{article}
\newcommand{\be}{\begin{equation}}
\newcommand{\ee}{\end{equation}}
\newcommand{\ba}{\begin{eqnarray}}
\newcommand{\ea}{\end{eqnarray}}
\newcommand{\p}{\partial}
\newcommand{\del}{\bigtriangledown}
\begin{document}
\begin{flushright}
\today
\end{flushright}
\vspace{.5in}
\begin{center}
{\Huge Chiral Perturbation on the Lightfront}\\
\vspace{.25in}
{\Large Alfred Tang}\\
\vspace{.15in}
{\em Max Planck Instit\"ut f\"ur Kernphysik}\\
{\em Postfach 103980, D-69120 Heidelberg, Germany.}
\end{center}
\vspace{5mm} \noindent
A new geometrical interpretation of chiral perturbation theory based
on a topological QCD model is explained in pictures.  This work is largely a
written summary of a talk presented at NAPP 2003, Drubovnik, Croatia.
\vspace{.3in}

\section{Introduction}
Chiral perturbation provides a framework to calculate corrections to nuclear
observables such as nucleon masses, charge radii and form factors.  When chiral
symmetry is broken, intermediate mesons are created as Goldstone bosons which
contribute to the tree level values of relevant physical observables.
Traditional chiral perturbation theory based on Feynman diagrams provides a
means to estimate the corrections that comes from the Goldstone bosons.
Unfortunately the calculations become intractable rapidly beyond the one loop
level.  The current doctrine of particle physics is a bottom-up approach that
tries to analyze the most fundamental units of nature in terms of elementary
particles with all the associating dynmanics and symmetries.  The bottom-up
approach of the particle picture (particularly in the framework of perturbative
QCD) cannot solve the nucleons because such systems are characterized by highly
non-linear many-body problems.  Recent developments of non-critical dimension
string theory in $AdS^5\times S_5$ with ${\mathcal N}=1$ shows some promises
in solving the nucleons by virtue of avoiding the the difficult perturbative
QCD calculations by effectively summing the infinite perturbative QCD series
on the string world sheets.  This approach is a paradigm shift from traditional
wisdom in a sense that the bottom-up approach of particle physics that begins
from elementary symmetries is replaced by the top-down approach of string
theory that start with global symmetries that hide the details on the particle
level.  Although supersymmetry is a beautiful symmetry, it may pose too many
constraints when modeling the nucleons.  The present work suggests an
alternative top-down approach by summing the infinite series on the string
worldsheets with no assumption of any supersymmetry nor extra dimensions.
This research is work in progress.  This paper explains the logic of this new
approach in picture format.  The materials shown here
are essentially the same as those presented in the talk given at NAPP 2003 in
Dubrovnik, Croatia, but with minor changes.  The introductory remarks of the
original talk, such as a brief review of chiral perturbation theory and
recent works related to $\chi$pt on the lightfront, are skipped in this paper
so that the main concept, {\em i.e.} topological QCD, is addressed
right away.

\section{Topological QCD}
\begin{figure}[h]
\begin{center}
\epsfig{file=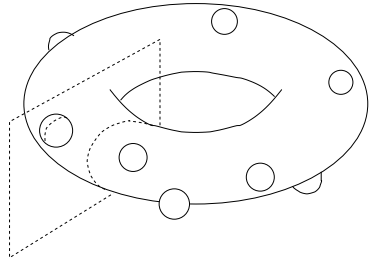,scale=.3}
\caption{The geometry of a nucleon in the instant form.  The torus represents
the nucleon and the spheres (gaugeballs) symbolize the vacuum.  The plane
is the surface of initial data that intersects the torus.  The instant form
Hamiltonian evolves plane of initial data along the torus in the time-like
direction and sweeps it through the gaugeballs so that the instant form
Hamiltonian sees a complicated vacuum.}
\label{instant}
\end{center}
\end{figure}

\begin{figure}[h]
\begin{center}
\epsfig{file=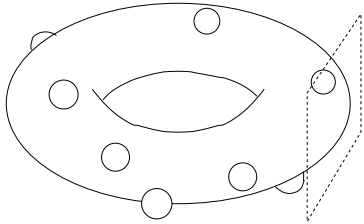,scale=.3}
\caption{The geometry of a nucleon in the front form.  The torus represents
the nucleon and the spheres (gaugeballs) symbolize the vacuum.  The plane
is the surface of initial data that is tangential to the torus.  The front form
Hamiltonian evolves the plane of initial data around the torus in a spiral
direction so that the front form Hamiltonian sees a simpler vacuum.}
\label{front}
\end{center}
\end{figure}

A geometrical interpretation of QCD is quite old.  For instance the gauge
field of a nucleon is usually represented as a torus $T^2$ in lattice QCD.  In
the same spirit, the disconnected graphs of the tadpoles can be modeled as
$S^2$ spheres.  A gauge field is represented by a worldsheet called the gauge
surface
from now on.  In the instant form, the geometry of the nucleon is represented
in Figure~\ref{instant}.  The plane of initial data intersects the gaugeballs
representing the vacuum.  The Hamiltonian evolves the plane of initial data
along the torus in the time-like direction.  In this case, the plane sweeps
through the gaugeballs so that the structure of the vacuum is not simple.
On the other hand, the initial data in the front form is parameterized along a
plane tangent to
the the torus as shown in Figure~\ref{front}.  The front form Hamiltonian
evolves the plane of initial data around the torus in spiral directions with
various winding modes.  Since the gaugeballs are disconnected from the torus,
the plane of initial data does not necessarily intersect the gaugeballs during
the evolution.  This picture recovers the typical result that lightfront QCD
is uniquely endowed with a simpler vacuum.

\begin{figure}[ht]
\begin{center}
\epsfig{file=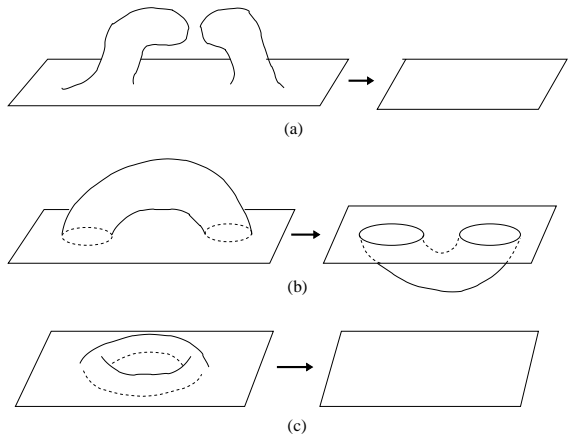,scale=.3}
\caption{Simplications that comes from topological equivalence. (a) A hairpin,
(b) a pion loop and (c) sea quarks.}
\label{topeq}
\end{center}
\end{figure}

At the end the worldsheets of the gauge fields can be summed using path
integrals.  The integrals are categorized according
to geni.  A genus 0 surface has no hole, a genus 1 surface has 1 hole and
so on.  All the topologically
equivalent surfaces are grouped together.  Topological equivalence greatly
simplifies calculations.  For instance, the geometry of all the hairpins and
sea quarks are equivalent to a plane and are effectively cancelled as shown in
Figures~\ref{topeq}(a) and (c).

\section{Non-Abelian Gauge as Torsion}
\begin{figure}[ht]
\begin{center}
\epsfig{file=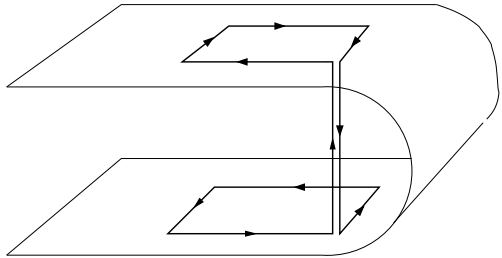,scale=.3}
\caption{The torsion vector from the loop on the upper surface is in the
opposite direction of that on the lower surface so that torsion is
effectively canceled in this topology.}
\label{abel}
\end{center}
\end{figure}

\begin{figure}[ht]
\begin{center}
\epsfig{file=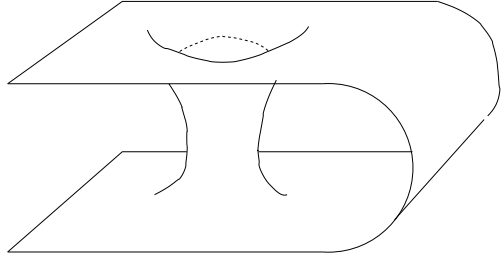,scale=.3}
\caption{A pion loop (shown as a wormhole) on a surface of torsion-free
toplogy.}
\label{local}
\end{center}
\end{figure}

In non-Abelian gauge theory, the partial derivative are modified by the
addition of an extra term that comes from minimal coupling as in
\be
D_\mu=\p_\mu+i\tau_c\,A^c_\mu.
\ee
With the extra term, the total derivatives no longer commute,
\be
D_\mu D_\nu-D_\nu D_\mu=\underbrace{\p_\mu \p_\nu-\p_\nu \p_\mu}_{0}
+i\tau_c\,F^c_{\mu\nu}.
\label{total}
\ee
The commutation of the total derivatives measures the shift of the gauge
field around a loop.  A non-zero shift indicates an implict twist in the gauge
configuration.  In general relativity, the torsion-free condition is written as
\be
\del_a\del_b f - \del_b\del_a f = 0,
\ee
where $f$ is a scalar field.  By analogy, we can interpret the right hand
side of Eq.~(\ref{total}) geometrically as torsion.  It is hoped that
this torsion can be gauged away by an appropriate choice of transformation.
Figure~\ref{abel} shows heuristically how torsion can be canceled with a
carefully chosen topology.
In general relativity, the mass of the field is related to curvature as in
\be
m\sim {1\over r},
\ee
so that the curvature of the edge of the folded surface is interpreted as
the effective gluon mass.  This interpretation is slightly different from that
given in the original talk at NAPP 2003 where the curvature was interpreted as
the mass of the constituent quark.  Figure~\ref{topeq}(b) suggests the
possibility that the pion loop can be constructed inside the gauge surface
as shown in Figure~\ref{local}.

\section{Global Torsion-free QCD Topology}
\begin{figure}[ht]
\begin{center}
\epsfig{file=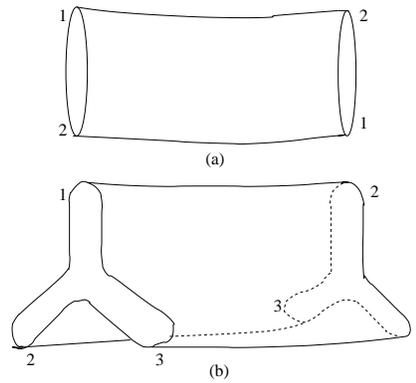,scale=.3}
\caption{The topological structures of the gauge fields of a meson and a
baryon.  The identifications of the boundaries are labeled.}
\label{hadron}
\end{center}
\end{figure}

\begin{figure}[ht]
\begin{center}
\epsfig{file=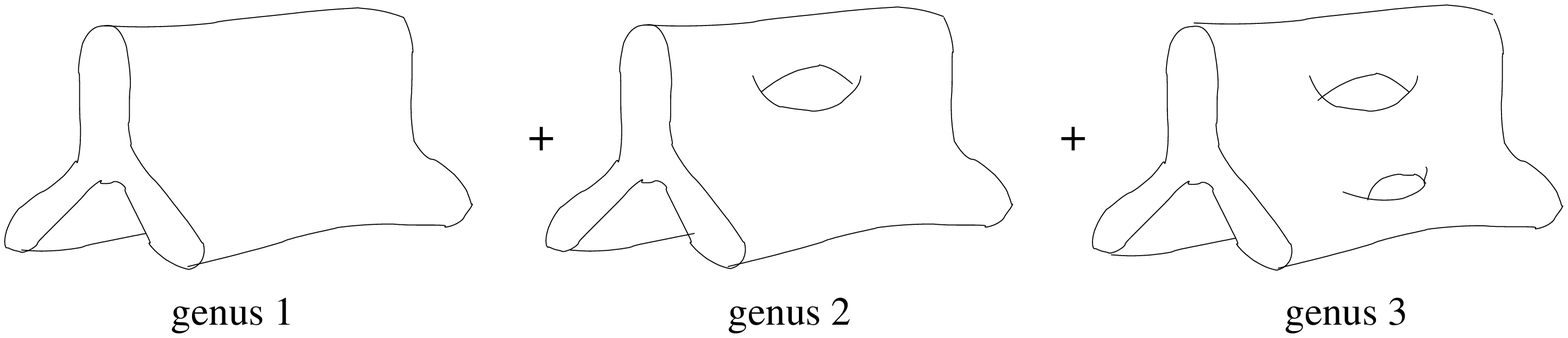,scale=.4}
\caption{Chiral perturbation on the lightfront is reduced to a sum of
graphs of various geni. }
\label{chiral}
\end{center}
\end{figure}

It is reasonable to assume that the presence of constituent quarks must affect
the geometry of the gauge field.  A good guess is that the global topology
of a meson has two edges
and that of a baryon has three.  Although torsion (the non-Abelian part of the
gauge theory) is gauged away by a suitable transform, it survives as a twist
on the transformed field configuration.  Figure~\ref{hadron} shows a sketch
of the topological structures of the gauge fields of a meson and a baryon.
The twist causes the topology of the gauge field to have an unusual
boundary condition similar to a M\"obius trip.  The topology of the
baryon has been called the 3-M\"obius trip and the ``triniton'' in previous
presentations.
In this picture, chiral perturbation on the lightfront is represented by
a sum of these graphs of different geni as shown in Figure~\ref{chiral}.

\section{Conclusion}
The present work illustrates the concept of topological QCD in pictures.
This research is work in progress.  Many claims made in this presentation
are based on intuition but are now being subjected to the rigor of
mathematics.  The contributions of the constituent quarks have not yet been
included in the pictures of this work.  These deficiencies will be amended in a
future publication.  Some colleagues have also pointed out that this research
should aim to produce relevant predictions--and rightly so.  It
suffices to say that topological QCD is suggestive so far and is hopeful as a
candidate to analyze the structure of the hadrons.  Currently a search is
underway to find a suitable transform that maps the non-Abelian gauge field
configuration to the torsion-free twisted $n$-M\"obius surface.  Then the
mathematics of the path integrals over the surfaces of the twisted tori
need to be formulated.  At last it is hoped that this model can explain
traditionally difficult theoretical problems such as confinement and asymptotic
freedom.  It is also hoped that new technology of non-perturbative QCD
calculations is discovered in the process.

\end{document}